\documentclass[12pt]{iopart}

\usepackage[colorlinks,linkcolor=blue,citecolor=blue,urlcolor=blue]{hyperref}
\usepackage{eqnarray}
\usepackage{graphicx}
\usepackage{upgreek}
\usepackage[dvipsnames]{xcolor}

\renewcommand{\vec}[1]{\mathbf{#1}}

\newcommand{\eqref}[1]{(\ref{#1})}

\begin{document}

\title{Functionalizing   Fe adatoms   on  Cu(001)  as a nanoelectromechanical system}

\author{Michael Sch\"uler$^1$, Levan Chotorlishvili$^1$, Marius Melz$^1$,
  Alexander Saletsky$^2$, Andrey Klavsyuk$^2$, Zaza Toklikishvili$^3$, Jamal Berakdar$^1$}
\address{$^1$ Institut f\"ur Physik, Martin-Luther-Universit\"at Halle-Wittenberg, 06099 Halle, Germany}
\address{$^2$ Faculty of Physics, Moscow State University, Moscow, 119991 Russia}
\address{$^3$ Department of Physics, Tbilisi State University, Chavchavadze Avenue 3, 0128 Tbilisi, Georgia}
\ead{jamal.berakdar@physik.uni-halle.de}

\begin{abstract}
  This study demonstrates how the spin quantum dynamics of a single Fe
  atom adsorbed on Cu(001) can be controlled and manipulated by the
  vibrations of a nearby copper tip attached to a nano cantilever by
  virtue of the dynamic magnetic anisotropy. The magnetic properties
  of the composite system are obtained from \emph{ab initio}
  calculations in completely relaxed geometries and turned out to be
  dependent considerably on the tip-iron distance that changes as the
  vibrations set in.  The level populations, the spin dynamics
  interrelation with the driving frequency, as well as quantum
  information related quantities are exposed and analyzed.
\end{abstract}

%\pacs{32.80.Rm, 33.20.Lg, 33.20.Wr, 81.05.ub}
\submitto{\NJP}
\maketitle
%\ioptwocol

%\section{Color coding}
%
%%----------------------------------------
%% Correspondence color coding of the changes/
%% author. Please remove before resubmission.
%%----------------------------------------
%
%\begin{itemize}
%  \item \ftbf{Levan}
%  \item \textcolor{red}{Andrey}
%  \item \textcolor{OliveGreen}{Michael}
%\end{itemize}

\section{Introduction}

Microelectromechanical (MEMS) or nanoelectromechanical systems (NEMS)
are at the verge of the classical-quantum world
\cite{chan_laser_2011,PhysRevA.82.012333,0295-5075-93-1-18003,PhysRevLett.103.063005}
and can thus sense, possibly coupled, quantum-classical properties.
For instance, tiny vibrating cantilever were shown to detect a single
spin \cite{rugar_single_2004}.  The sensitivity depends on the mean
phonon number with the cantilever dynamics turning quantum as the
phonons number decreases.  Related to these observations, this field
promises a new rout to quantum information nanomechanical devices. An
example is the setup consisting of a single nitrogen-vacancy (NV)
centre in a diamond nanocrystal deposited at the extremity of a SiC
nanowire \cite{arcizet_single_2011}. The quantum NV spin dynamics is
observed to be coupled to the nanomechanical oscillator by means of
the time-resolved nanocrystal fluorescence and photon-correlation
measurements. This dynamic can be influenced by external fields such
as a non-homogeneous magnetic field.  A clear advantage of utilizing
the spin-degrees of freedom of the NV is their long decoherence times
even at room temperatures \cite{PhysRevB.79.041302, Treutlein,
  PhysRevA.90.033817, PhysRevLett.92.076401,
  Awschalom,PhysRevB.88.085201}.  Further phenomena emerge when
considering strongly coupled nonlinear NEMS in which case phenomena
such as nonlinear resonances can be exploited for the control of the
energy transfer between the coupled NEMS
\cite{PhysRevB.79.165309,Chotorlishvili_JPB,Lifshitz_2003,Lifshitz_2007}.

In the present work we propose a new type of NEMS based on a single
magnetic Fe adatom deposited on a Cu(001) substrate. A proper choice
of the driving frequency allows controlling the level populations in
the system.
The proposed (scanning-tunneling microscopy) STM-type or (atomic force
microscopy) AFM-type setup is thus a hybrid system utilizing the
quantum nature of single adsorbed atoms or molecules on the surface,
which were designed as studied in an impressively controllable way
experimentally (for example in
Refs. \cite{loth_measurement_2010,loth_bistability_2012,schwobel_real-space_2012}).
{The magnetic properties of Fe and Co adatoms on a
  Cu$_2$N/Cu(100)-c$(2\times 2)$ surface were determined experimentally
  via  x-ray magnetic dichroism measurements \cite{PhysRevB.92.184406}. } 

The possibly classical cantilever dissipative dynamics is coupled the
quantum spin dynamics of the adsorbates since, as demonstrated below,
the magnetic anisotropy is affected by the tip-adsorbate distance, and
hence by the tip vibrational motion. This coupling might be exploited
to access the topology or the local magnetic properties of spin
systems \cite{wolter_spin_2012}.

\begin{figure}[b]
  \centering
  \includegraphics[width=0.6\textwidth]{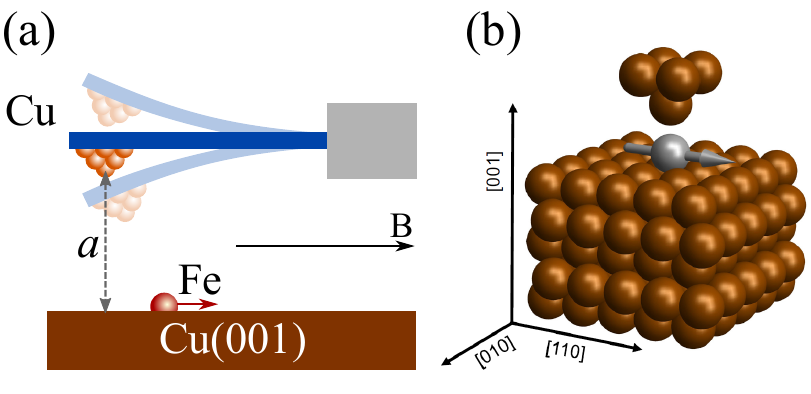}
  \caption{(a) The prototypical model system for coupling the spin
    dynamics of a single magnetic Fe adatom on the Cu(001) substrate
    to the mechanical vibrations of a nano-sized cantilever. (b) The
    cluster used to represent the copper substrate, the copper tip and
    the Fe adatom. \label{fig1}}
\end{figure}

\section{Theoretical framework}

Specifically, we  consider a single magnetic Fe atom deposited
on a Cu(001) surface. A similar setup consisting of  Fe or Mn on copper
coated by an $\rm{Cu}_2$N overlayer,
 was shown to have
 a large magnetic anisotropy and relaxation
times\cite{hirjibehedin_large_2007, loth_measurement_2010}. Our
calculations are carried out in the presence of a tip apex as in AFM
experiments and reveal a substantial dependence of the magnetic
anisotropy on the distance between the tip and the Fe adatom.  Thus, in the
 proposed setup the magnetic properties of the single-atom
are coupled to the oscillations of a nano-sized cantilever carrying
the tip apex (FIG.~\ref{fig1}(a)). As the characteristic frequencies
of such nano-mechanical oscillators are known to reach the gigahertz
regime\cite{agarwal_amplitude_2011}, frequencies in the range of
$\sim$100~GHz become feasible upon a further downscaling of the
cantilever and can thus match the typical energy scale of the spin
system (which is in the range of few meV).

%The paper is organized as follows: In the section II we present results of
%the ab initio calculations, in the section III we discuss classical
%spin dynamics, and in the last section IV we investigate quantized
%oscillations of the cantilever.

\section{\emph{ab initio} calculations}

\textit{Ab initio} density-functional calculations of the ground state
and the energy difference upon changing the magnetization axis of the
Fe atom were performed using the projector augmented-wave (PAW)
technique~\cite{PhysRevB.50.17953,PhysRevB.59.1758}, as implemented in
the Vienna \textit{ab initio} simulation package
(VASP)~\cite{PhysRevB.48.13115}.  The calculations are based on
density-functional theory with the generalized gradient approximation
(GGA)~\cite{PhysRevB.44.13298,PhysRevB.45.13244}.  We used the same
methodology used in previous calculations of the magnetic anisotropy
of Co and Fe adatoms on Rh(111), Pt(111) and Cu(100)
substrates~\cite{PhysRevB.81.104426,SurfSci.612.48,JETPL.99.646}.
{We note in this context that  STM experiments were performed after the 
	tip were in  contact with the surface, and hence the tip is most likely covered by the surface
  material~\cite{RevModPhys.75.1287}.  In this case  a tip with
  surface atoms is often used in computer
  calculations~\cite{PhysRevLett.83.2765,PhysRevB.73.153404}.}
  
  Our computational models consists 125 copper atoms representing the
surface, the iron atom, and 5 additional atoms for simulating the
presence of a tip apex, as depicted in FIG.~\ref{fig1}(b).  The unit
cell has a size of 12.87~\AA \ in the $x$ and $y$ directions (parallel
to the surface), whereas the extent of the $z$ direction
(perpendicular to the surface) amounts to 31.89~\AA. At this slab
thickness, the interaction between the tip and the repeated image of
the surface is negligible. A cutoff energy of 300 eV is used. The
calculations including spin-orbit coupling require a fine k-point mesh
for the Brillouin-zone integrations. Test calculations were performed
for iron atom on a Cu(001) surface for three different
\textit{k}-point grids: $3\times3\times1$ , $3\times3\times2$ , and
$5\times5\times1$ generated by the Monkhorst-Pack
scheme~\cite{PhysRevB.13.5188}, in conjunction with a modest Gaussian
smearing method. A $3\times3\times1$ grid provided the best compromise
between accuracy and computational efforts.

The calculations were performed in two steps. First the coordinates of
the iron atom and the positions of the atoms in the three topmost
layers of the substrate (apart from the tip) were optimized using
scalar-relativistic calculations until the forces on all unconstrained
atoms were converged to less than 0.01~eV/\AA.  In the second step,
the geometry and the electronic and magnetic ground states resulting from
the scalar-relativistic calculations were used to initialize the
relativistic calculations including spin-orbital coupling.  Recent
work~\cite{JOPC_42_426001} demonstrated that relaxations of Fe and Co
adatom on Pt(111) with and without spin-orbit coupling are almost
identical.

\begin{figure}[ht!]
  \centering
  \includegraphics[width=\textwidth]{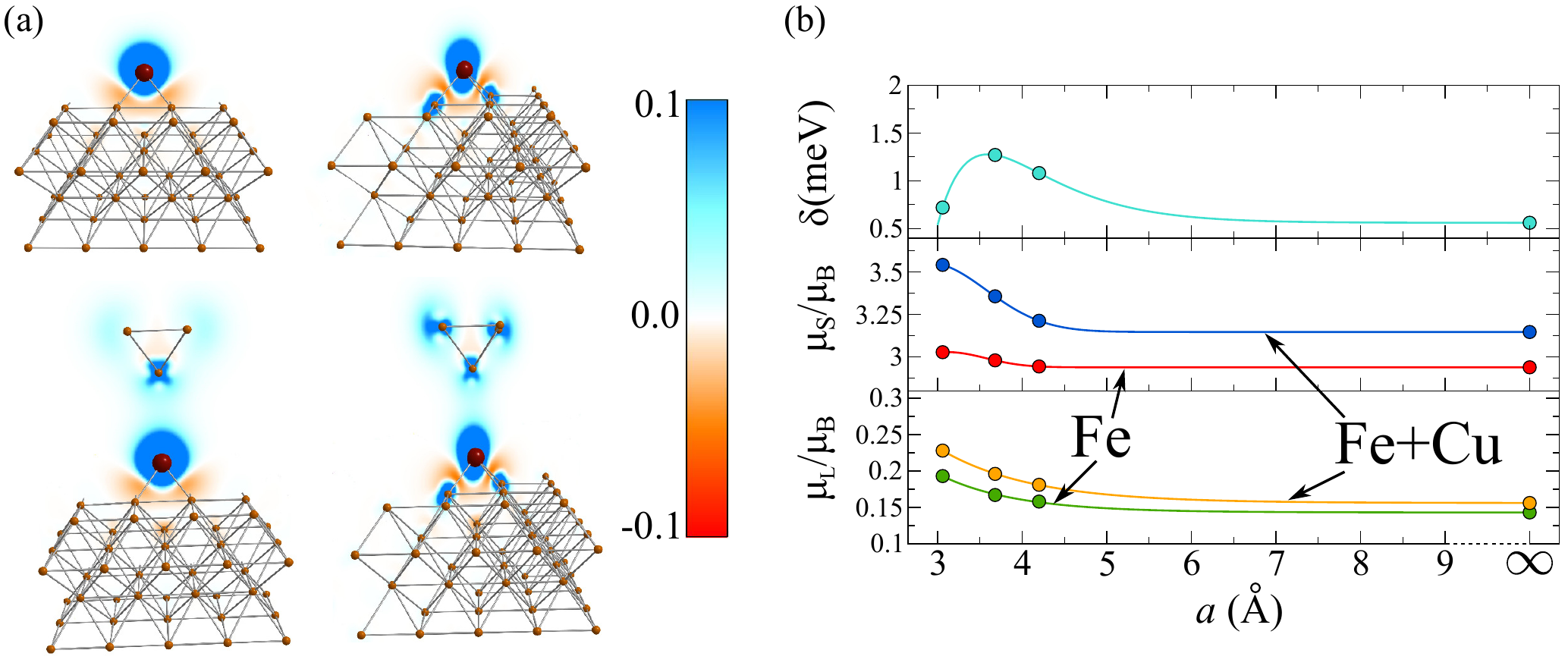}
  \caption{(a) The spin density of Cu-Fe system without tip (upper two
    figures), and with the tip situated 4.5~\AA \ above the adsorbate
    (lower two figures). We show cuts of the spin density (unit:
    \AA$^{-3}$) along the planes passing through the Fe atom $(0,1,0)$
    (upper and lower left) and $(1,1,0)$ (upper and lower right
    figures). The number of atoms shown were reduced to emphasize the
    $\rm{C}_\mathrm{4v}$ symmetry.  (b) The prefactor of the
    anisotropy (upper panel), the magnetic moment originating from the
    spin (middle panel), and from the orbital moment (lower
    panel). {A positive sign of the magnetic
      anisotropy parameter corresponds to a perpendicular easy axis. } 
    The values from our \emph{ab initio} calculations (dots) are well
    described by Gaussian-type fitting functions (full lines).  The
    red and the green curves correspond to the values when integrating
    the spin density over a small sphere around the Fe atom, whereas
    the neighboring Cu atoms have been also included for the blue and
    the orange curves.
    \label{fig2}}
\end{figure}

After a geometry optimization of the full cluster (apart
from the tip) for every position of the tip, we computed the magnetic
anisotropy energy as the difference of the respective ground state
energies upon varying the magnetization axis. We found that the
dependence on the angle $\theta$ measured from the Fe-tip axis is well
described by the lowest-order anisotropy term $\delta \sin^2 \theta$,
as it is known for similar systems. The dependence on the angle $\phi$
measured along the plane on the other hand turned out to be rather
weak. Furthermore, we  analyzed the spin density $n_\uparrow(\vec
r) - n_\downarrow(\vec r)$ to investigate the degree of
localization of the magnetization. The result is presented in
FIG.~\ref{fig2} for small values of the density in two characteristic
planes along the symmetry directions. We conclude that the Fe atom
slightly polarizes the tip and the substrate below. Especially for the
latter we observe the typical behavior of a spin density associated
with this kind of anisotropy.

However, the major contribution to the magnetization is confined
within the direct vicinity of the Fe atom,
confirming that the effective surface spin can be
interpreted as the magnetic moment of  few atoms.

\section{Modeling the spin dynamics}

In FIG.~\ref{fig2}(b) we show our results for the dependence of the
magnetic anisotropy parameter $\delta$ and of the magnetic moments
associated with the spin ($\mu_\mathrm{S}$), and the angular momentum
($\mu_\mathrm{L}$) for four different values of the distance $a$
between the last tip atom and the iron atom.  The spin magnetic moment
on the iron atom without the tip-adatom interaction is 2.94
$\mu_\mathrm{B}$. This result agrees well with previous density
functional calculations \cite{PhysRevB.68.205422}.  It should be
noted, that the magnetic anisotropy parameter $\delta$ for an atom on
the surface is very sensitive to the interatomic distances
\cite{PhysRevB.68.104410,PhysRevB.70.224419} and the arrangement of
the atom \cite{PhysRevB.81.104426,PhysRevLett.102.257203}. It was
demonstrated that the structural relaxation of the adatom and the
substrate reduces significantly the magnetic anisotropy energy
\cite{PhysRevB.68.104410,PhysRevB.70.224419}.  Therefore, compared to
\textit{ab initio} calculations for the Fe adatom on the ideal Cu(001)
surface, our magnetic anisotropy energy obtained in a fully relaxed
geometry is several times less than the value presented in
Ref.~\cite{PhysRevB.68.024433}.

Based on the fitting functions displayed in FIG.~\ref{fig2}(b) we are now
able to formulate the Hamiltonian describing the effective surface
angular momentum with the parametric dependence on $a$ as
\begin{equation}
  \label{eq:ham1}
  \hat{H}(a) = -\big[g_S(a) + g_L(a)\big]\mu_\mathrm{B} B_0 \hat{J}_x - \delta(a) \hat{J}^2_z \ .
\end{equation}
We assume a magnetic field with a strength $B_0$ is applied along the
$x$ axis. Approximating the total angular momentum with 2 turns out to
be an adequate description and can be confirmed experimentally by
means of inelastic tunneling
spectroscopy~\cite{hirjibehedin_large_2007,loth_measurement_2010}.
{The   distance-dependent gyromagnetic ratios for spin (angular) momentum
  $g_S(a)$ ($g_L(a)$) account for the varying magnitude of the total
  magnetic moment (see Fig.~\ref{fig2}(b)), as extracted from our
  \emph{ab initio} calculations.}
We fix $B_0$ to the value of 4~T
and take  the ground state as the initial state. As one can
readily show for eq.~\eqref{eq:ham1}, the expectation value with
respect to all eigenstates of $\hat{J}_y$ and $\hat{J}_z$ is exactly zero. This
 holds true even for the case of the time-dependent Schr\"odinger
equation, when replacing $a\rightarrow a(t)$.

\begin{figure*}[ht!]
  \centering
  \includegraphics[width=0.9\textwidth]{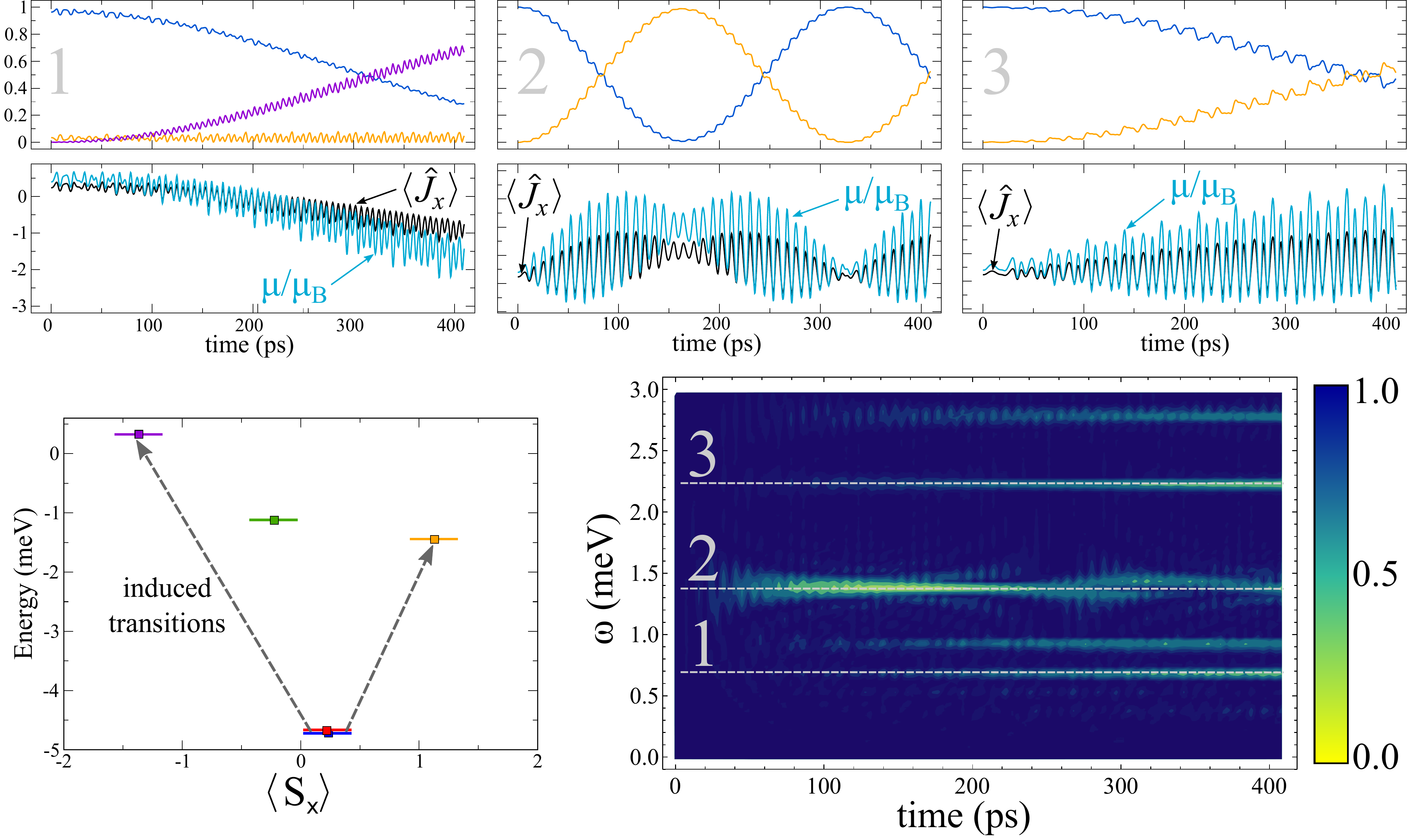}
  \caption{The population dynamics arising due to the periodic driving
    $a(t)$. The lower panels show the initial configuration (left), and
    the probability to measure the ground state as a function of the
    time and the driving frequency $\omega$ (right). The grey dashed
    lines with their respective labels 1,2 and 3 denote three selected
    frequencies. The corresponding population and the spin dynamics are
    displayed in the upper three panels.\label{fig3}}
\end{figure*}

To map out the spin dynamics for a representative case we choose $a$
by $a_0=4~\AA$ \ and $B_0=4$~T. The energies of the eigenstates
$|\xi_n\rangle$ (on the ordinate axis) and the expectation values of
$J_x$ (abscissa) are shown in the inset in FIG.~\ref{fig3}. For a high
density of phonons the spin dynamics originates from an oscillation of
the tip apex according to $a(t)=a_0 + b\sin(\omega t)$ 
{for $t>0$ (we
assume $a(t)=a_0$ for $t\le 0$). This corresponds to a setup where the
system is initially in its ground state and is driven out of
equilibrium by the cantilever oscillations for $t>0$. The oscillation
amplitude is chosen as $b=0.9$~\AA.}

{Before discussing the results, let us elaborate on the 
	qualitative aspects of the dynamics. Since the spin is
  driven by an effectively time-dependent anisotropy, i.\,e. the
  coupling to the operator $\hat{J}^2_z$, the induced transitions
  allow for changing the spin projection $\langle \hat{J}_x \rangle$
  only. The magnetic moment $\mu$ will thus remain parallel to
  $\langle \hat{J}_x \rangle$. Therefore, only the longitudinal spin
  dynamics can be induced, limiting the transitions from the ground
  state to only the two excited states that match in symmetry.}

\begin{figure}[t]
  \centering
  \includegraphics[width=0.9\columnwidth]{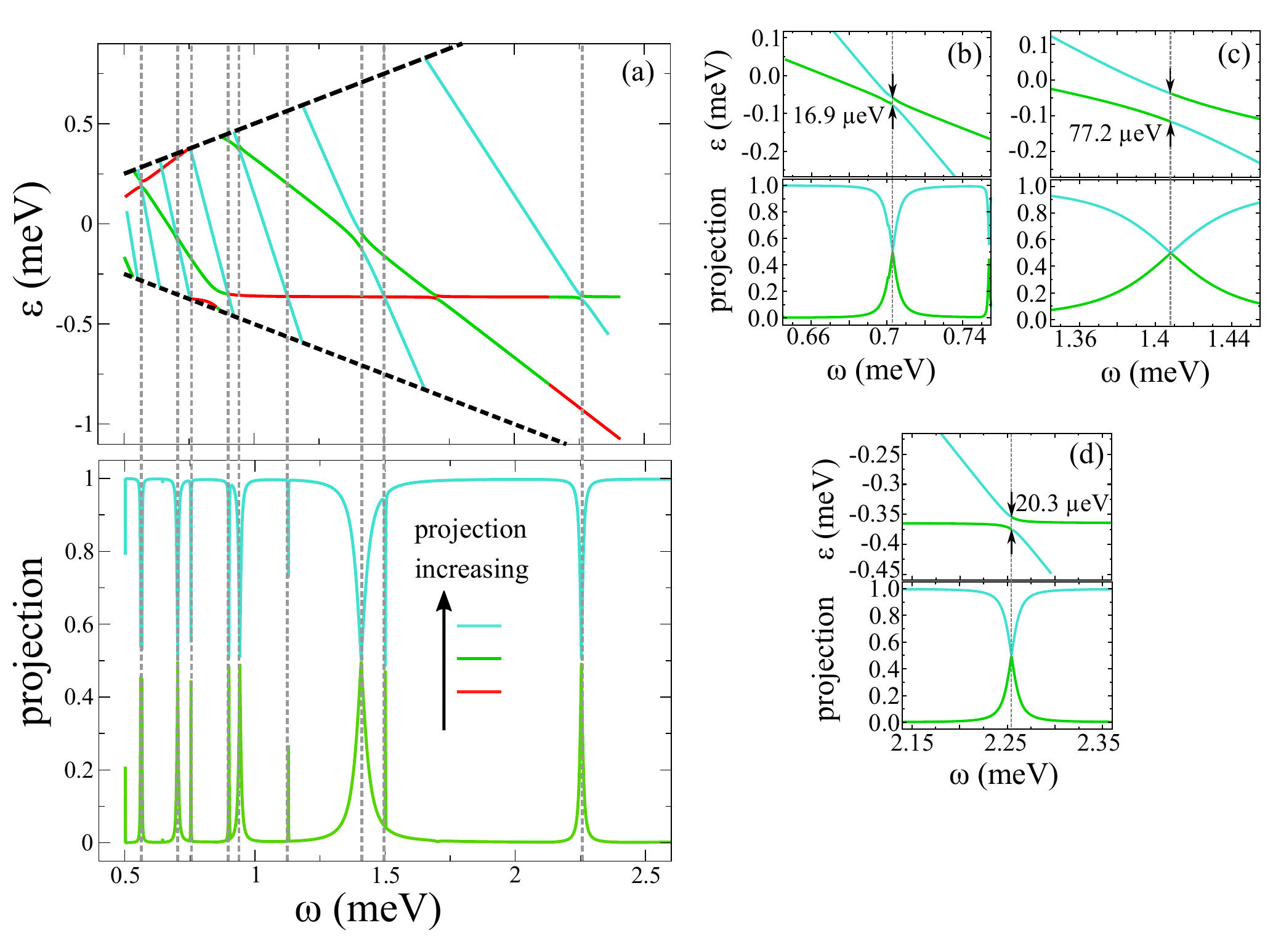}
  \caption{(a) Upper panel: the quasienergies $\varepsilon_n$ (folded
    back to the first Brioullin zone, indicated by the black dashed
    line) as a function of the driving frequency $\omega$. The color
    coding corresponds to the projection of the eigenvectors
    $|f_n(0)\rangle$ onto the initial state,
    $|\langle \xi(0)|f_n(0)\rangle|^2$, which is displayed in the
    lower panel. The vertical grey lines connect the points where the
    projection approaches $1/2$ with the avoided crossing points. (b)
    A zoom around $\omega = 0.704$~meV , (c)
    $\omega=1.408$~meV and (d)$ \omega=2.254$~meV. The gap width of
    the crossings is given in the inset. \label{fig4}}
\end{figure}

In FIG.~\ref{fig3} we present the resulting spin dynamics in
dependence on $\omega$. The color map plot (lower right figure) shows
the population of the ground state (which we have chosen as the
initial state). Interestingly, the magnetic moment is hardly affected
by the variation of the magnetic anisotropy for the major part of the
frequency range. Apart from that, a couple of distinct lines indicate
an optimal setting for the parameters  to drive the angular
momentum to some excited states. A more detailed analysis reveals that
the dynamics for the frequencies indicated by the dashed lines
(labelled by 1,2,3) exhibits almost Rabi-like transitions from the ground state
to a single excited state. This behavior can  be explored further by a
Floquet analysis. We therefore expand the angular momentum wave
function as
\begin{equation}
  \label{eqfloquet}
  |\xi(t)\rangle=\sum^{2J+1}_{n=1} c_n |\phi_n(t) \rangle \
= \sum^{2J+1}_{n=1} c_n e^{i \varepsilon_n t}|f_n(t) \rangle \ ,
\end{equation}
where $\varepsilon_n$ are the quasi energies and $|\phi_n(t)\rangle
=e^{i \varepsilon_n t}|f_n(t)\rangle$ the Floquet eigenvectors. Both
can be obtained by solving the eigenvalue problem of the
time-evolution operator $U(t,0)$ at $t=T\equiv 2\pi/\omega$, since
\begin{equation}
  \label{eqfloqueteigen}
  \hat{U}(T,0)|f_n(0)\rangle = e^{i \varepsilon_n T}|f_n(0)\rangle
\end{equation}
(note that $|f_n(t+T)\rangle = |f_n(t) \rangle$). Before discussing
the solution of eq.~\eqref{eqfloqueteigen}, let us briefly revisit how
the dynamics shown in FIG.~\ref{fig3} can be explained within the
Floquet theory. Eq.~\eqref{eqfloquet} expresses  the expansion of the
time-dependent spin state in terms of the orthonormal Floquet states
$|f_n(t)\rangle$, with the projection coefficients $c_n$. For the case
$c_n=c_{n_0}\delta_{n,n_0}$, the projection $\langle \xi(0)|\xi(t)
\rangle$ amounts to $\exp[-i \varepsilon_{n_0} t] \langle
f_{n_0}(0)|f_{n_0}(t) \rangle$, such that the population of the
initial state remains one at multiples of $T$. Assuming on the other
hand $c_n= (\delta_{n,n_1}\pm\delta_{n,n_2})/\sqrt{2}$ yields the
stroboscopic time evolution
\begin{equation}
\label{eqscenario2}
|\langle \xi(0)|\xi(k T)
\rangle|^2 = \cos^2[(\varepsilon_{n_1}-\varepsilon_{n_2})k T].
\end{equation}
These two scenarios explain the dynamics observed in FIG.~\ref{fig3},
where slow, Rabi-like population transfer (with a frequency
corresponding to the difference of two quasienergies) is overlayed
with fast oscillations (which originate from the overlaps of the type
$\langle f_n(0)|f_m(t) \rangle$ and are thus periodic with the
frequency $\omega$). The quasienergies obtained from
eq.~\eqref{eqfloqueteigen} are presented in FIG.~\ref{fig4}(a) (upper
panel), along with the projection $|c_n|^2=|\langle \xi(0)|f_n(0)
\rangle|^2$ (lower panel). As pointed out, the decisive  factor for the
depletion of the ground state is at least two projection coefficients
being different from zero. For this reason, we have ordered the
quasienergies according to the magnitude of $|c_n|^2$. As it turned
out, only two of the Floquet states display a significant contribution
to the initial state. Therefore, only their projection is shown in
FIG.~\ref{fig4}(a). For the complete picture of the behaviour of the
quasienergies however, the third state and its eigenvalue are included in
the upper panel.

The vertical lines in FIG.~\ref{fig4}(a) demonstrate that the scenario for the
dynamics according to eq.~\eqref{eqscenario2}  occurs only at the
crossing points of the quasienergies, where (at least) two branches
exchange their character, that is the magnitude of their projection
coefficients. A more detailed analysis reveals that all
crossings are avoided crossings. The difference of the quasienergies
 becomes thus relatively small, leading to the slow dynamics in
FIG.~\ref{fig3}. For the exemplary values of $\omega$,
 FIG.~\ref{fig4}(b)--(c) provides a magnification of the crossing
points and gives their width. Converting the quasienergy gap into a
time scale results in exactly the characteristic time of the slow
dynamics in FIG.~\ref{fig3}.

\section{Entanglement measures}

With cooling down the system, oscillations of the nanocantilever
become inherently quantum. Thus the nanocantilever can detect
inter-level transitions of the single spin. This statement is generic
for a broad class of the nanomechanical systems and is valid for our
model as well. Our model is exactly solvable (which is in fact its
merit) and allow  to
explore  analytically the entanglement between the cantilever and the system. In contrast to
the above, where we investigated the case of a moderate elongations of
the cantilever, leading to a nonlinear coupling, we consider linear
coupling only. In addition to the feasibility  of  analytical
solutions, the quantum fluctuations and the oscillations of the cantilever occur
on a smaller scale, while large elongations are associated with the classical
case which is studied above.

{In order to quantify the entanglement in the system, we explore
  von Neumann entropy. In the quantum information theory  the von Neumann
  entropy is known as the "entanglement entropy" of the reduced density
  matrix. The technical details of the von Neumann entropy and hence of the
   reduced density matrix in our case are given in Appendix A.}

To construct a quantized model, the tip-substrate distance is replaced
by $a \rightarrow a_0 + \Delta a \sum_{\alpha}(\hat{a}_\alpha + \hat{a}^\dagger_\alpha) \ $,
%\begin{align}
%  a \rightarrow a_0 + \Delta a \sum_{\alpha}(\hat{a}_\alpha + \hat{a}^\dagger_\alpha) \ ,
%\end{align}
where $\hat{a}^\dagger_\alpha$ ($\hat{a}_\alpha$) is the
creation (annihilation) operator of the cantilever modes. Modern
technologies enabled the fabrication of dual mode ($\alpha=1,2$)
cantilevers, for more details see \cite{Solares1,Solares2}.

The resulting Hamiltonian, in lowest order in the oscillation
amplitude $\Delta a$ reads
\begin{equation}
  \label{eq:Hquant}
  \hat{H} = -\left[g_S(a_0)+g_L(a_0)\right]\mu_\mathrm{B}B_0 \hat{J}_x
  + \sum_\alpha \Omega_\alpha \hat{a}^\dagger_\alpha\hat{a}_\alpha
\nonumber - \Delta a\delta'(a_0) \hat{J}^2_z \sum_{\alpha=1,2}(\hat{a}_\alpha + \hat{a}^\dagger_\alpha) \ .
\end{equation}
Similar to the classical case analyzed above, the transition operator
$\hat{J}^2_z$ allows the transition from the ground state to two
excited states only. Hence, the Hamiltonian~\eqref{eq:Hquant} can be
reduced to a three-level system in spin space. We assume that the
cantilever frequencies $\Omega_{1,2}$ match the excitation energies
$\omega_{1,2}=E_{1,2}-E_0$.

We solve directly analytically for the Schr{\"o}dinger equation
\begin{equation}\label{6}
  i \frac{\partial |\Psi \rangle}{\partial t} = \hat{H} |\Psi \rangle,
\end{equation}
using the following ansatz
\begin{eqnarray}
\label{7}
|\Psi(t)\rangle &= C_1(t)|n_1,n_2,1 \rangle + C_2(t)|n_1-1,n_2,2 \rangle + C_3(t)|n_1,n_2-1,3 \rangle \\ & \quad + C_4(t)|n_1,n_2,4 \rangle + C_5(t)|n_1,n_2,5 \rangle \nonumber.
\end{eqnarray}
Here $n_1,n_2$ quantify the number of phonons in the cavity with the
frequencies $\Omega_{1},\Omega_{2}$.  Taking into account
Eq. \eqref{6}, \eqref{7} we consider the resonance condition
$E_2-E_1\approx \Omega_1$,
$E_3-E_1\approx \Omega_2$. After standard calculations we
obtain

\begin{eqnarray}
\label{8}
  C_1(t)=&\exp(-i\Delta_1 t) \bigg\lbrace C_1(0)\cos(\gamma\sqrt{n_1+n_2}t)\nonumber
\\ &-i\frac{C_2(0) \sqrt{n_1} \sin(\gamma \sqrt{n_1+n_2} t)}{\sqrt{n_1 + n_2}}\nonumber
\\ &-i\frac{C_3(0) \sqrt{n_2} \sin(\gamma\sqrt{n_1 + n_2}t)}{\sqrt{n_1+n_2}}  \bigg\rbrace, \\
 \label{9}
C_2(t)=&\exp(-i(\Delta_2-\Omega_1) t) \bigg\lbrace -i\frac{C_1(0) \sqrt{n_1}
 \sin(\gamma \sqrt{n_1+n_2} t)}{\sqrt{n_1 + n_2}}\nonumber
\\ &+\frac{C_2(0) \left( n_1 \cos(\gamma\sqrt{n_1 + n_2}t)+n_2 \right)}{n_1+n_2}\nonumber
 \\ &+ C_3(0)\frac{\sqrt{n_1n_2}}{n_1+n_2} \left( \cos(\gamma \sqrt{n_1 + n_2}t)-1 \right) \bigg\rbrace,
\end{eqnarray}

% \begin{equation}\label{8}
% \begin{aligned}
% C_1(t)=&\exp(-i\Delta_1 t) \bigg\lbrace C_1(0)\cos(\gamma\sqrt{n_1+n_2}t)\\&-i\frac{C_2(0) \sqrt{n_1} \sin(\gamma \sqrt{n_1+n_2} t)}{\sqrt{n_1 + n_2}}\\&-i\frac{C_3(0) \sqrt{n_2} \sin(\gamma\sqrt{n_1 + n_2}t)}{\sqrt{n_1+n_2}}  \bigg\rbrace, \\
% C_2(t)=&\exp(-i(\Delta_2-\Omega_1) t) \bigg\lbrace -i\frac{C_1(0) \sqrt{n_1} \sin(\gamma \sqrt{n_1+n_2} t)}{\sqrt{n_1 + n_2}}\\&+\frac{C_2(0) \left( n_1 \cos(\gamma\sqrt{n_1 + n_2}t)+n_2 \right)}{n_1+n_2} \\&+ C_3(0)\frac{\sqrt{n_1n_2}}{n_1+n_2} \left( \cos(\gamma \sqrt{n_1 + n_2}t)-1 \right) \bigg\rbrace, \\
% C_3(t)=&\exp(-i(\Delta_3-\Omega_2) t) \bigg\lbrace -i\frac{C_1(0) \sqrt{n_2} \sin(\gamma \sqrt{n_1+n_2} t)}{\sqrt{n_1+n_2}}\\&+ C_2(0)\frac{\sqrt{n_1n_2}}{n_1+n_2} \left( \cos(\gamma \sqrt{n_1 + n_2}t)-1 \right)\\&+\frac{C_3(0) \left( n_2 \cos(\gamma\sqrt{n_1 + n_2}t)+n_1  \right)}{n_1+n_2}  \bigg\rbrace.
% \end{aligned}
% \end{equation}
 In Eq. \eqref{8} and \eqref{9} we introduced the notation
$\Delta_m=E_m+n_1\Omega_1+n_2\Omega_2$.   Further simplifications
enabling an analytical treatment are
$\langle 2| \hat{J}^2_z|1\rangle \approx \langle 3|
\hat{J}^2_z|1\rangle \equiv g_0$  giving rise to the effective
coupling constant $\gamma = g_0\Delta a \delta'(a_0)$.

{While the solution Eq. \eqref{7}-\eqref{9} is obtained for a
  fixed value of the magnetic field $B_{0}=4$T, it is valid for an
  arbitrary field. } 
{Changing the magnetic field
     rescales the level spacing, thus leading to a slight
    rescaling of the Rabi-like transition frequencies.}

The quantities we are interested in, such as level populations
$I_n(t)=C_n(t)C_n^*(t)$ and von Neumann entropy
$S=-tr\big(\hat{\rho}\ln\hat{\rho}\big)$ (where
$\hat{\rho}=\sum_{m n}C_n(t)C_m^*(t)|m\rangle\langle n|$ is the density matrix of the system) can
be calculated directly from \eqref{8} and \eqref{9}.  For this purpose
we need to consider the averaging of Eq. \eqref{8}--\eqref{9} over the
phonon distribution functions for the coherent states
$w_{n_{1,2}}=\frac{\sqrt{\lambda_{1,2}^{n_{1,2}}}}{\sqrt{n_{1,2}!}}\exp\left(\frac{-\lambda_{1,2}}{2}\right)$.
Here $\lambda_{1,2}$ is the mean phonon number $\lambda_{1,2} \gg 1$
corresponding to the classical limit.

\begin{figure}[t]
  \centering
  \includegraphics[width=\textwidth]{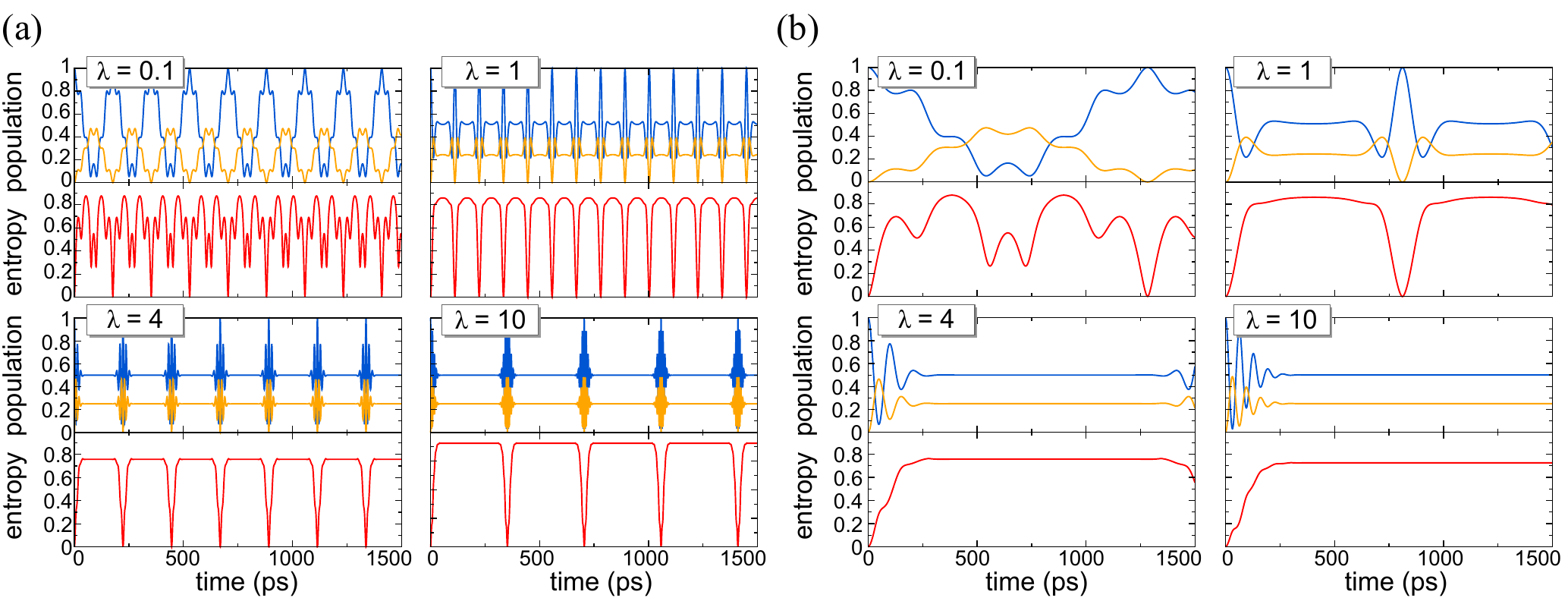}
  \caption{Population dynamics (averaged over the phonon number) of
    the ground state (blue) and excited state (orange) along with the
    von Neuman entropy for different mean photon numbers $\lambda$. In
    (a) the tip-sample distance is fixed at
    $a_0=3.0$~\AA\ , while in (b) $a_0=4.0$~\AA.\label{fig5}}
\end{figure}

For the calculations of the average level populations we perform
summation over the phonon numbers $n_{1}$ and $n_{2}$:
\begin{equation}
  \overline{I_{n}(t)}=\overline{|C_{n}(t)|^{2}}=\sum_{n_1,n_2=0}^{\infty}w_{n_1}^{2}w_{n_2}^{2}|C_{n}(t)|^{2}
=\sum_{n_1,n_2=0}^{\infty}\frac{e^{-\lambda_1}e^{-\lambda_2}\lambda_1^{n_1}\lambda_2^{n_2}}{n_{1}!n_{2}!}|C_{n}(t)|^{2} \ .
\end{equation}
The coefficients $C_{n}(t)$ have a sharp maximum near the mean phonon
numbers $\lambda_{1,2}\gg 1$, and the width $\Delta n_{1,2}$ of their
distribution is rather small $\Delta n_{1,2} \gg \lambda_{1,2}$. This
allows performing the summation analytically and obtaining expressions
for the level populations and for the von Neumann entropy. The
explicit expression is presented in \ref{app:entropy}.

With the analytical solutions at hand, we can now study the population
dynamics and the von Neumann entropy of the spin system due to the
interaction with the cantilever. FIG.~\ref{fig5} depicts this dynamics
with an oscillation amplitude of $\Delta a = 0.1$~\AA. Due to the
strong dependence of the anisotropy $\delta(a)$ on the tip-sample
distance, the dynamics at $a_0=3.0$~\AA\ (FIG.~\ref{fig5}(a)) and
$a_0=4.0$~\AA\ (FIG.~\ref{fig5}(b)) occurs on very different time
scales.  We clearly see that quantum revivals in level populations are
synchronized with the sudden death of von Neumann entropy. This
behavior is more prominent in the case of a strong coupling
(FIG.~\ref{fig5}(a)). Obviously with the increase of the phonon number
$\lambda$ the period of quantum revivals becomes larger. In the limit
of the classical field $\lambda\gg 1$ the revival time tends to
infinity. Meaning that the classical field like thermal bath
thermalizes the system and leads to irreversibility.

\section{Conclusions}

In summary, we performed \emph{ab initio} calculations of the magnetic
properties of a single Fe atom adsorbed on Cu(001).  We demonstrated
that the electronic and the magnetic properties of adatoms are strongly
affected by the tip-surface distance.  Based on these results
we  proposed a new  type of NEMS consisting of  a
 single magnetic Fe adatoms deposited  on a Cu(001) substrate and analyzed
its fundamental properties and possible operation scheme.

\appendix

\section{von Neumann entropy\label{app:entropy}}

Using Eq.~\eqref{8}--\eqref{9}, a straightforward derivation yields
the von Neumann entropy
\begin{eqnarray}
 && S=-\eta_{1}\ln\eta_{1}-\eta_{2}\ln\eta_{2}-\eta_{3}\ln\eta_{3}. \nonumber\\
 &&\lambda_{1}=\lambda_{2}=\lambda,\beta=\gamma\sqrt{2\lambda}t,\alpha=\frac{\gamma t}{\sqrt{2 \lambda}},\nonumber \\
 &&\langle I_{1}(t)\rangle=\frac{1}{2}\Big(1+\exp\big[2\lambda\big(\cos\alpha-1\big)\big]\cos\big(\beta+2\lambda\sin\alpha\big)\Big);\nonumber\\
 &&\langle I_{2}(t)\rangle=\langle I_{3}(t)\rangle=\\
 &&=\frac{1}{4}\Big(1-\exp\big[2\lambda\big(\cos\alpha-1\big)\big]\cos\big(\beta+2\lambda\sin\alpha\big)\Big);\nonumber\\
 &&\langle I_{4}(t)\rangle=\langle I_{5}(t)\rangle=0 \ . \nonumber
\end{eqnarray}
Here, we used the following notation in order to obtain a compact
expression:
 \begin{eqnarray}
 &&a=\frac{1}{2}\Big(1+\exp\big[2
 \lambda\big(\cos\alpha-1\big)\big]\cos\big(\beta+2\lambda\sin\alpha\big)\Big);\nonumber\\
 &&b=\frac{\sqrt{2}}{4}\exp\big[2
 \lambda\big(\cos\alpha-1\big)\big]\sin\big(\beta+2\lambda\sin\alpha\big);\nonumber\\
 &&d=\frac{1}{4}\Big(1-\frac{1}{4\lambda}\Big)\Big(1+\exp\big[2
 \lambda\big(\cos\alpha-1\big)\big]\cos\big(\beta+2\lambda\sin\alpha\big)\Big);\nonumber\\
 &&\eta_{1}=\frac{1}{16\lambda}\Big(1-\exp\big[2\lambda\big(\cos\alpha-1\big)\big]\cos\big(\beta+2\lambda\sin\alpha\big)\Big);\\
&&\eta_{2}=\frac{1}{4}\big(1+2d+a+\sqrt{9a^2+32b^2-6a(1+2d)+(1+2d)^2}\big);\nonumber \\
&&\eta_{3}=\frac{1}{4}\big(1+2d+a-\sqrt{9a^2+32b^2-6a(1+2d)+(1+2d)^2}\big);\nonumber
 \end{eqnarray}

\section*{Acknowledgments}

AK  acknowledges the financial support by the joint program of
MSU-DAAD Vladimir Vernadsky (A/12/89268) and Volnoe Delo foundation . Computational resources were provided
by the Supercomputing Center of Lomonosov Moscow State University.
The work was partially funded by the German Science Foundation under SFB 762.

%%%%%%%%%%%%%%%%%%%%%%%%%%%%%%%%%%%%%%%%%%%%%%%%%%%%%%%%%%%%%%%%%%%%
%                          references
%%%%%%%%%%%%%%%%%%%%%%%%%%%%%%%%%%%%%%%%%%%%%%%%%%%%%%%%%%%%%%%%%%%%

\section*{References}

\bibliography{lib1}{}
\bibliographystyle{iopart-num}

\end{document}